\documentclass[3p,times,procedia]{elsarticle}
\usepackage{nupha_ecrc}

\volume{00}

\firstpage{1}

\journalname{Nuclear Physics A}

\runauth{Md. Nasim (for the STAR collaboration)}

\jid{nupha}

\jnltitlelogo{Nuclear Physics A}

\usepackage{amssymb}
\usepackage[percent]{overpic}

\usepackage[figuresright]{rotating}

\begin{document}

\begin{frontmatter}



\dochead{}

\title{Measurements of $D_{S}^{\pm}$-meson production in Au+Au collisions at $\sqrt{s_{NN}}$ = 200 GeV in STAR}


\author{Md. Nasim (for the STAR collaboration)}

\address{University of California, Los Angeles, California-90095, USA}

\begin{abstract}
We present the first measurement of the nuclear modification
factor $R_{AA}$ and  elliptic flow $v_{2}$ of  $D_{s}$ in Au+Au
collisions at $\sqrt{s_{NN}}$ = 200 GeV with the STAR detector.
These results have been compared with those of other open charm mesons and strange mesons to determine
how the (possibly) strangeness equilibrated partonic matter affects the $D_{s}$ meson production. We find that the nuclear modification factor of $D_{S}$ are
systematically higher than unity and $D^{0}$ $R_{\rm{AA}}$. 
The ratio $D_{s}/D^{0}$ is shown as a function of transverse momentum for the
10-40$\%$ most central Au+Au collisions and compared with that in $p+p$  collisions obtained from PYTHIA.
It is also compared with that in Pb-Pb collisions at 2.76 TeV by the ALICE experiment. Our measurement indicates a hint of enhancement of
$D_{S}$ production in Au+Au collisions with respect to $p+p$ collisions as compared to non-strange $D$ mesons.
\end{abstract}

\begin{keyword}
strangeness, charm, quark-gluon plasma


\end{keyword}

\end{frontmatter}


\section{Introduction}
\label{intro}
Heavy quarks are considered an excellent probe for the
early dynamics in  heavy-ion collisions. They are produced on a short time scale ($\sim$0.08 fm/c for $c\bar{c}$ production) in
hard partonic scatterings during the early stages of 
nucleus-nucleus collisions. The probability of thermal production of
heavy-quark pairs in the high temperature quark-gluon plasma is expected to
be small. Among all  open charm
mesons, $D_{s}^{+}(c\bar s)$ and $D_{s}^{-}(\bar c s)$ mesons play a unique role to quantify
heavy quark diffusion and hadronization in heavy-ion collisions,
because of their valence quark compositions.
Recent theoretical model calculation predict that the nuclear modification factor ($R_{AA}$) of  $D_{S}$, when compared to those of non-strange $D$ mesons,
exhibits the influence of charm-quark recombination with thermal
partons through the enhanced production of strange quarks in the
deconfined matter~\cite{ds_prl}. Like multi-strange hadrons,  $D_{s}$ mesons are expected to freeze out early and have smaller hadronic interaction cross-sections compared to the non-strange $D$ mesons. Therefore, the elliptic flow ($v_{2}$) of
$D_{s}$  is considered to be a better measure of the partonic
contribution to the charm hadron $v_{2}$ than that of $D^{0}$ or
$D^{\pm}$~\cite{ds_prl}. The new Heavy Flavor Tracker (HFT) detector, which has been recently installed in the STAR experiment, provides a unique opportunity to reconstruct $D_{s}$ via displaced vertices at RHIC in Au+Au collisions at $\sqrt{s_{NN}}$ = 200 GeV.

\section{Data sets and analysis details }
\label{method}
\begin{figure}[!ht]
\begin{center}
\begin{overpic}[scale=0.35]{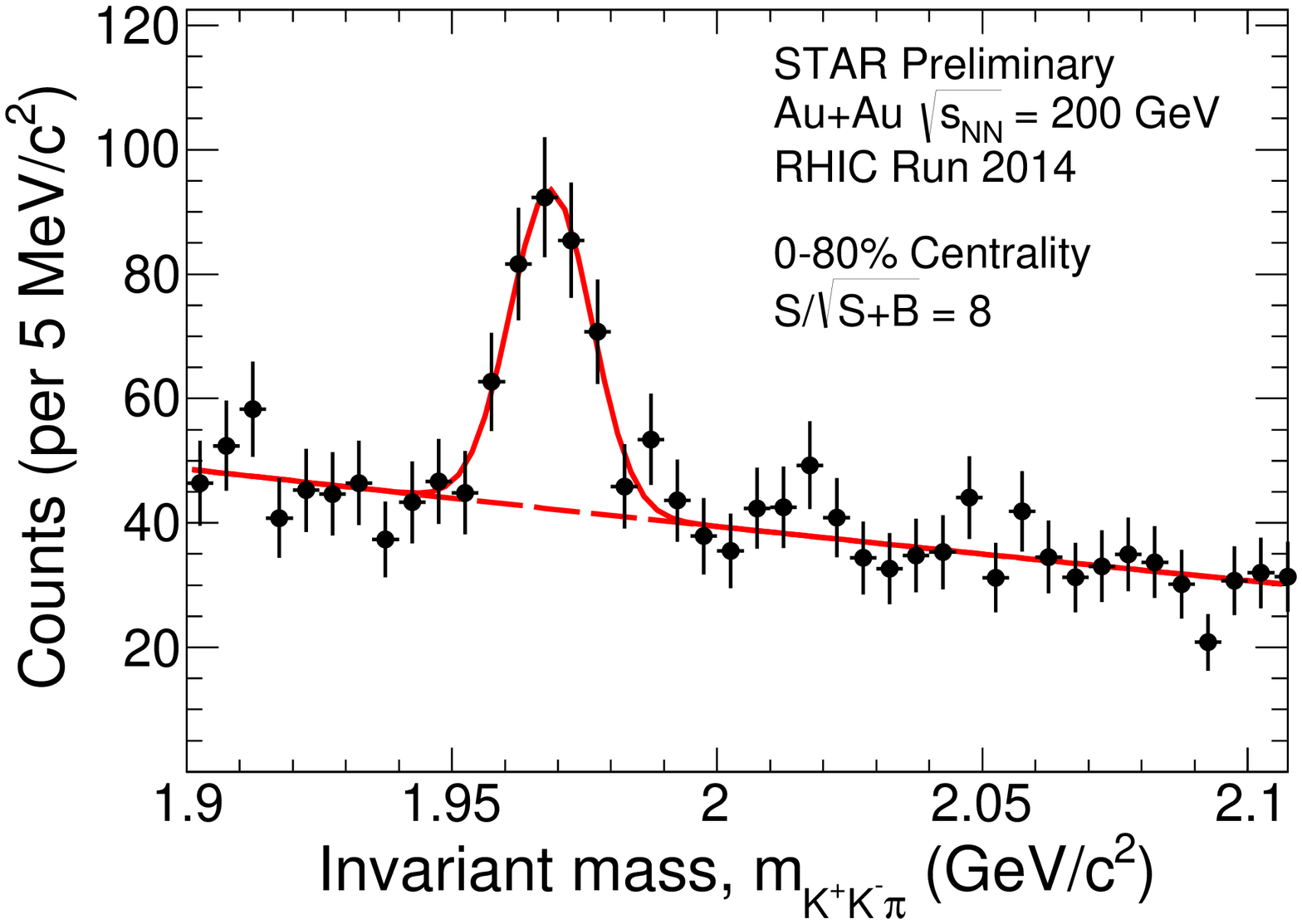}
\put (19,17) {\small 1.5 $< $ $p_{T}$ $<$ 5.0 GeV/$c$}
\put (17,65) {\small $(a)$}
\end{overpic}
\begin{overpic}[scale=0.355]{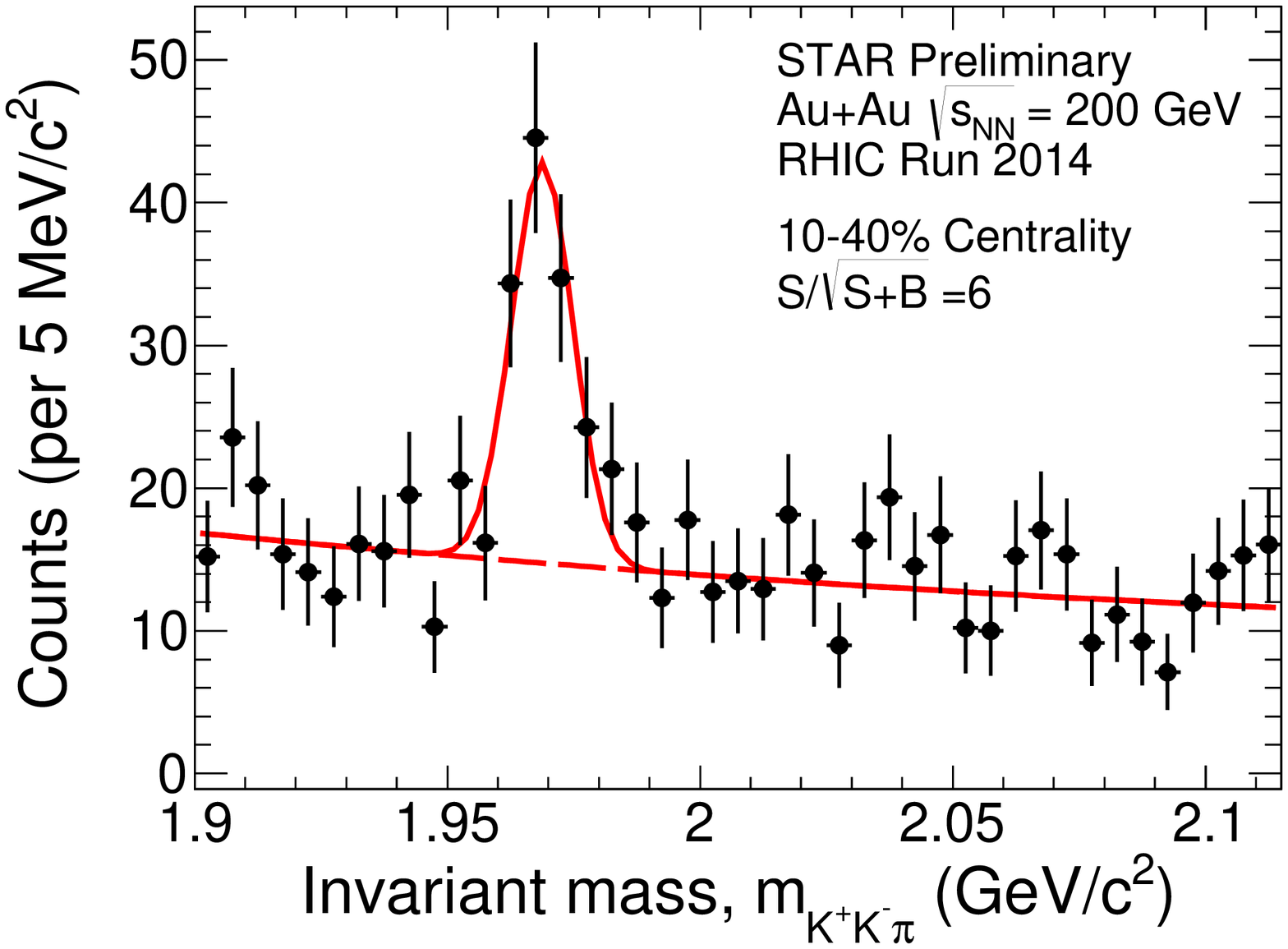}
\put (21,17) {\small 2.5 $< $ $p_{T}$ $<$ 5.0 GeV/$c$}
\put (21,65) {\small $(b)$}
\end{overpic}
\caption{(Color online)  The invariant mass distributions for $D_{S}$
  at  $\sqrt{s_{NN}}$ = 200 GeV for (a) 0-80$\%$ centrality, 1.5 $<$ $p_{T}$ $<$
  5.0 GeV/$c$ and (b) 10-40$\%$ centrality, 2.5 $<$ $p_{T}$ $<$
  5.0 GeV/$c$.}
\label{sig_08}
\end{center}
\end{figure}

The results presented here are based on an analysis of about 750 million
minimum bias events taken during the 2014 Au+Au run at
$\sqrt{s_{NN}}$= 200 GeV.
The Time Projection Chamber (TPC) and Time-of-Flight (TOF) detectors
with full azimuth coverage are used for particle identification in the
central rapidity ($\it{y}$) region ($|\it{y}|<$ 1.0).                                                                                                       
We reconstruct $D_{S}$  through the decay channel, $D_{S}^{\pm}$  $\longrightarrow$ $\phi$ ($\phi$ $\longrightarrow$ $\it{K}^{+}$ + $\it{K}^{-}$) + $\pi^{\pm}$.
Topological and kinematic cuts are applied to reduce the combinatorial background.
A first order polynomial function is used to describe 
the combinatorial background and a Gaussian function for the signal peak as                                                                                      
shown in Fig.~\ref{sig_08}.
The HFT detector is used to reconstruct the decay                                                                                                
vertices.  The HFT detector is made of  three  layers, named as PiXeL
detector (PXL),  Intermediate Silicon Tracker (IST) and Silicon 
Strip Detector (SSD)~\cite{contin}.  The state-of-the-art thin 
Monolithic Active Pixel Sensors (MAPS) technology has been used in the PXL. There are  two layers of MAPS  in the PXL,
which are placed at radii of 2.8 and 8 cm from the centre of the beam pipe, respectively.
The track pointing resolution of the HFT detector is about 46 $\mu$m for 750 MeV/c kaons.\\
\section{Results}
\label{results}
\begin{figure}[!ht]
\begin{center}
\begin{overpic}[scale=0.34]{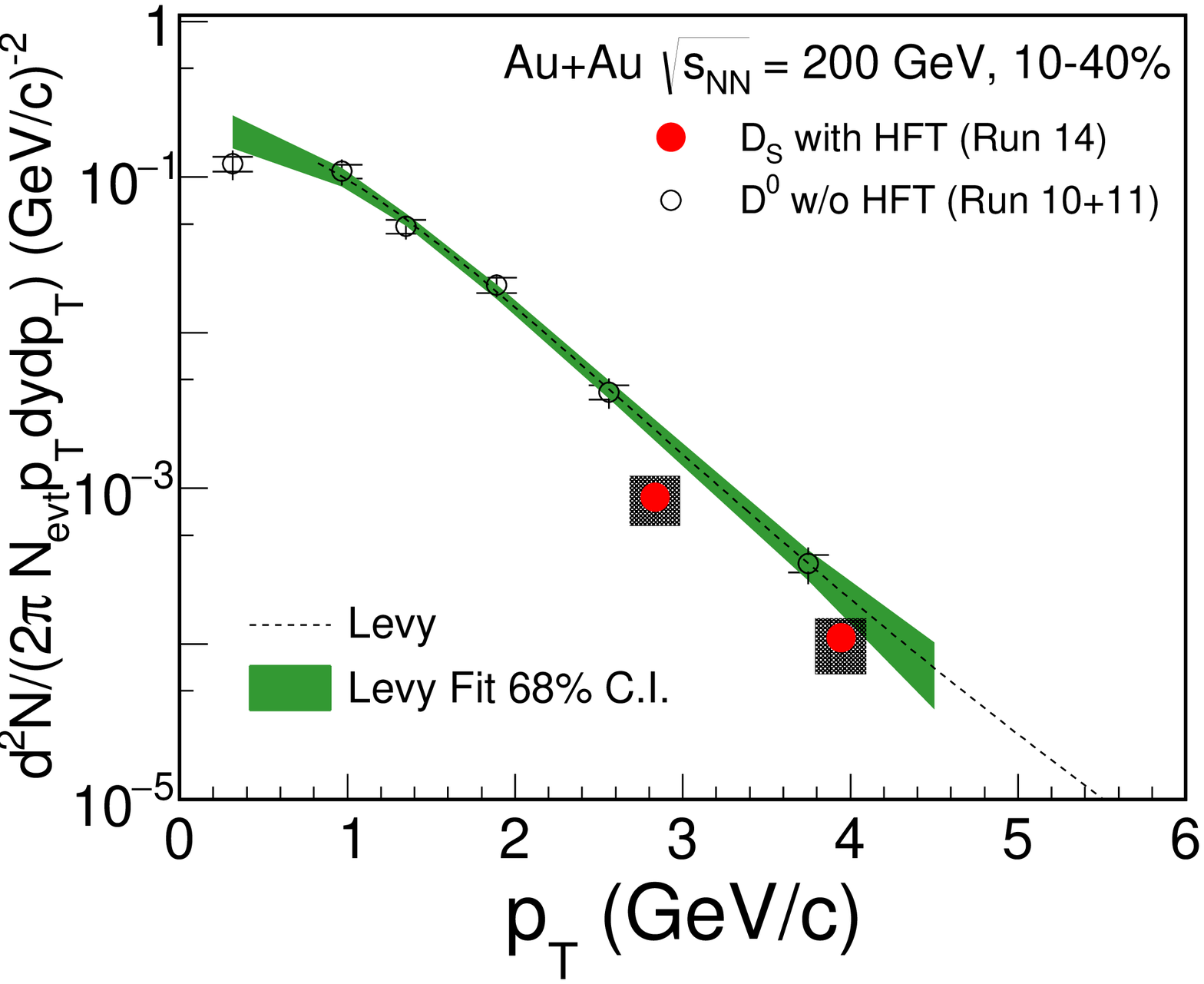}
\put (60,55) {\small \it STAR Preliminary}
\end{overpic}
\begin{overpic}[scale=0.34]{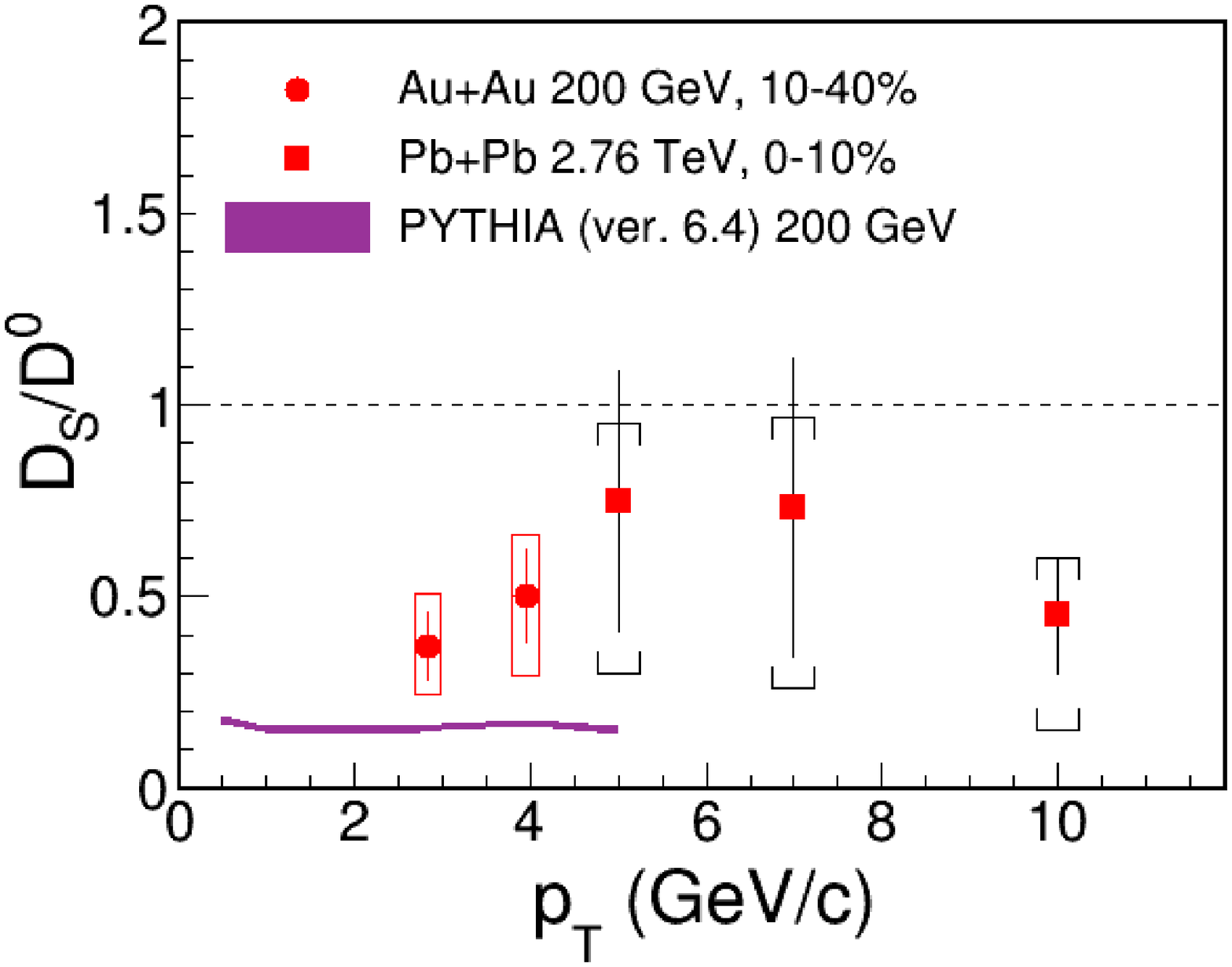}
\put (19,52) {\small \it STAR Preliminary}
\end{overpic}
\caption{(Color online) Left panel: The invariant yield of $D_{S}$  and
$D^{0}$ as a function of $p_{T}$ in Au+Au collisions at $\sqrt{s_{NN}}$= 200 GeV  for 10-40$\%$
centrality. Systematic uncertainties are shown by solid box and cap
symbol on $D_{S}$ and $D^{0}$, respectively. Vertical lines represent
the statistical errors. Right panel: The ratio $D_{S}/D^{0}$ as a
function of $p_{T}$. Solid red circles represent the data in Au+Au
collisions at $\sqrt{s_{NN}}$= 200 GeV  for 10-40$\%$, while solid
squares represent the results in Pb+Pb collisions at  $\sqrt{s_{NN}}$=
2.76 TeV  for 0-10$\%$. Open boxes and cap symbols are the systematic
error on corresponding data points. Magenta bands correspond to
$D_{S}/D^{0}$ in $p+p$ collisions at $\sqrt{s_{NN}}$= 200 GeV  calculated using PYTHIA (ver.6.4) model. }
\label{ds_by_d0}
\end{center}
\end{figure}

\begin{figure}[!ht]
\begin{center}
\begin{overpic}[scale=0.35]{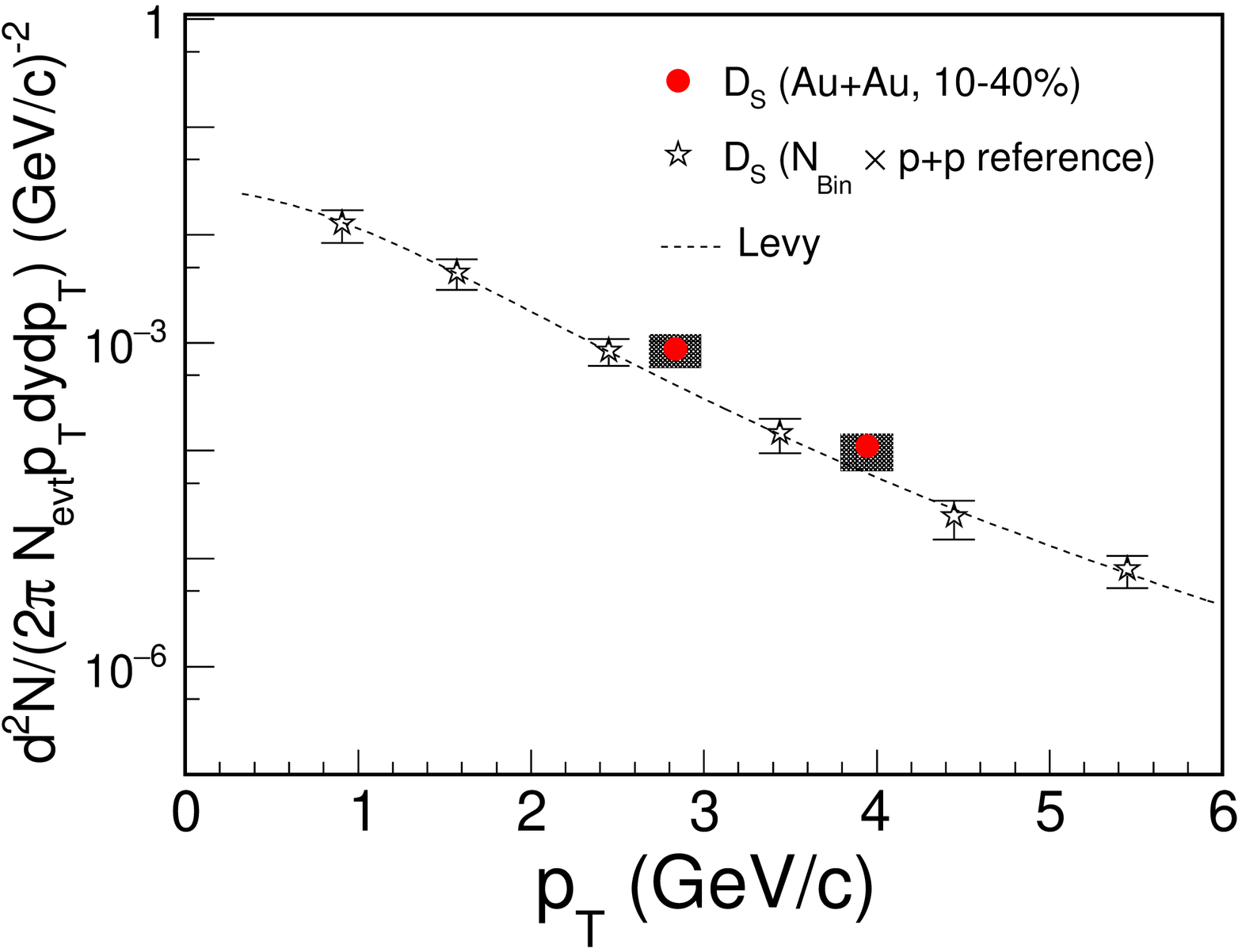}
\put (25,25) {\small \it STAR Preliminary}
\end{overpic}
\begin{overpic}[scale=0.33]{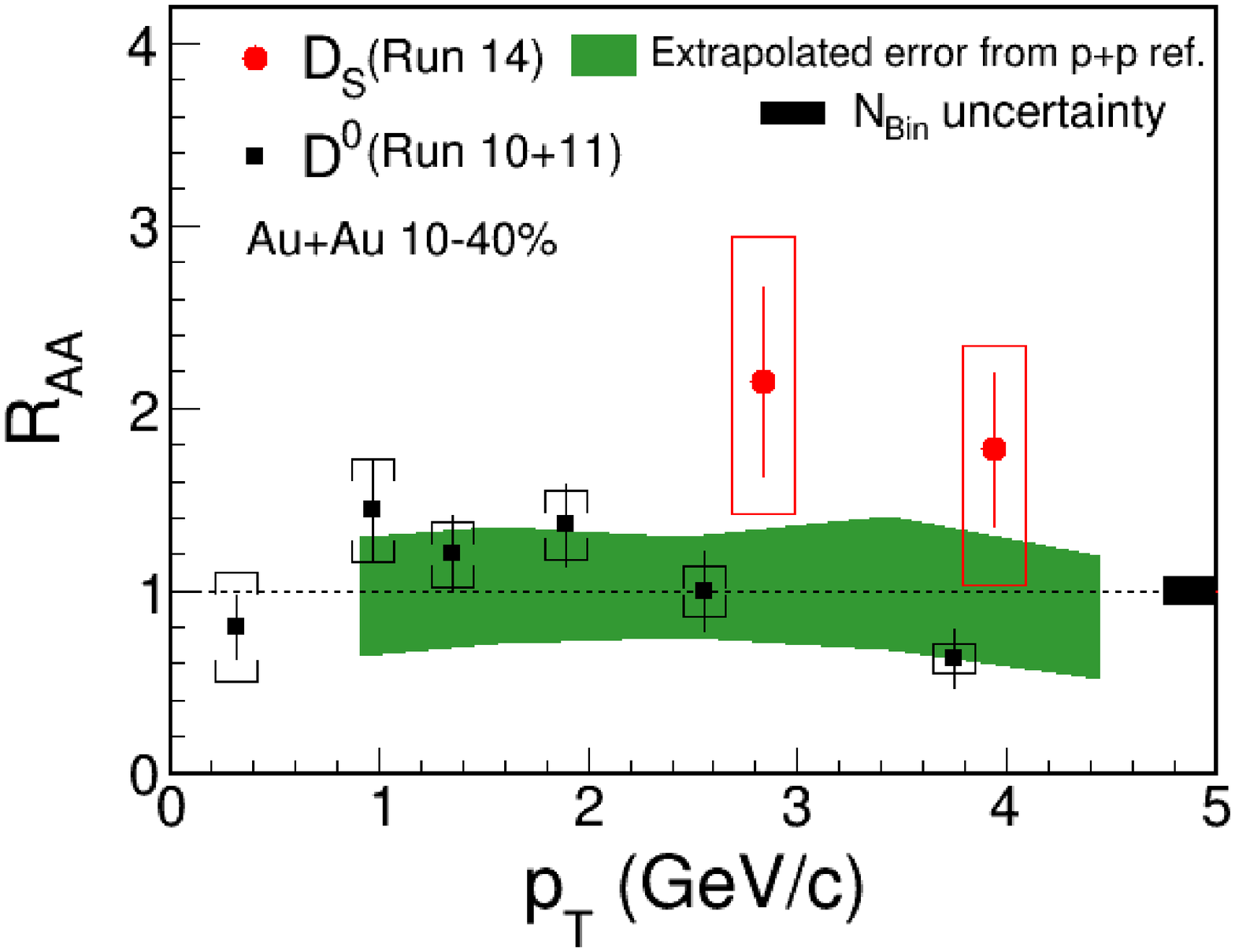}
\put (20,50) {\small \it STAR Preliminary}
\end{overpic}
\caption{(Color online) Left panel: The invariant yield of $D_{S}$ as a function of $p_{T}$ in  Au+Au collisions 
for 10-40$\%$ centrality (red filled circle) and in $p+p$ collisions (open black star) which is obtained from measured charm
cross-section in $p+p$ collisions at $\sqrt{s}$ = 200 GeV~\cite{stard0pp}. 
Right panel:  The nuclear modification factor $R_{\rm{AA}}$ of  $D_{S}$ at mid-rapidity ($|y| < 1.0$) in
Au+Au collisions at $\sqrt{s_{NN}}$ = 200 GeV for 10-40$\%$
centrality.  Open boxes and cap symbols are the systematic
error on corresponding data points. Statistical errors are shown by
vertical lines. Green bands represents the extrapolated error from
reference $D_{S}$ spectra in $p+p$ collisions.}
\label{raa}
\end{center}
\end{figure}

\begin{figure}[!ht]
\begin{center}
\begin{overpic}[scale=0.355]{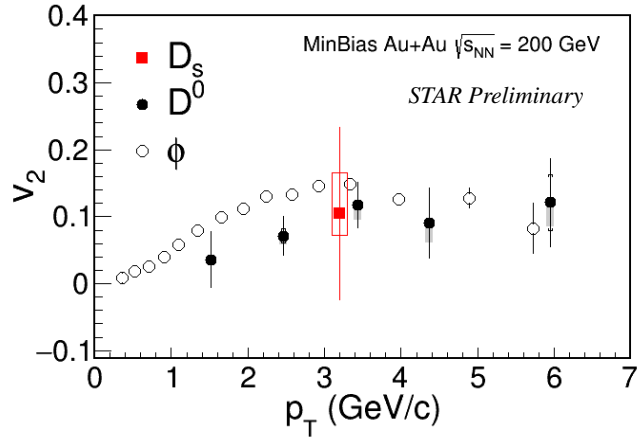}
\put (65,50) {\small \it STAR Preliminary}
\end{overpic}
\caption{(Color online) Elliptic flow as a function of
$p_{T}$ for $D_{S}$, $D^{0}$~\cite{d0v2} and $\phi$~\cite{200GeV_run11_star} for 0-80$\%$ centrality in Au + Au collisions at
$\sqrt{s_{NN}}$ = 200 GeV.  Open boxes and cap symbols are the systematic
error on corresponding data points. Shaded grey band on $D^{0}$
$v_{2}$ is for non-flow estimation. Vertical lines are statistical errors. }
\label{v2_fig}
\end{center}
\end{figure}

The invariant yield of $D_{S}$ (average yield of $D_{S}^{+}$ and $D_{S}^{-}$)  as a function of
transverse momentum ($p_{ T}$) for the 10-40$\%$ most 
central Au+Au collisions at $\sqrt{s_{NN}}$= 200 GeV is shown in the left panel of Fig.~\ref{ds_by_d0}. The solid
boxes are the systematic errors on $D_{S}$ yield and statistical
errors are shown by vertical lines. The systematic errors have been
calculated by using different techniques for background subtraction
and varying topology cuts. The $D_{S}$  yield is corrected for its reconstruction efficiency. The TPC
efficiency is calculated by embedding simulated tracks into real event
background. The efficiency related to the HFT detector and topology cuts has been calculated  from $D_{S}$ decay
simulation based on HFT  to TPC track matching ratio and track pointing
resolution directly from data. The invariant yield of $D^{0}$  for 10-40$\%$ centrality, published by STAR~\cite{stard0}, is also
shown for comparison in the left panel of Fig.~\ref{ds_by_d0}. To calculate the ratio between the yields of $D_{S}$
and $D^{0}$, we have fitted the $D^{0}$ spectra by a Levy function as
shown by the dashed line in the left panel of Fig.~\ref{ds_by_d0}. The ratio,
$D_{S}/D^{0}$, as a function of $p_{T}$ is shown in the right panel of
Fig.~\ref{ds_by_d0}. To compare with the $D_{S}/D^{0}$ ratio in $p+p$ collisions, we have used the result from the PYTHIA 6.4 Monte Carlo generator~\cite{pythia}, which is shown as a
magenta band. We can see from Fig.~\ref{ds_by_d0} that the ratio of
$D_{S}/D^{0}$ in Au+Au collisions at $\sqrt{s_{NN}}$= 200 GeV is less than unity and systematically higher than that in $p+p$
collisions from PYTHIA. We have also compared our results with  the ALICE
measurement in Pb+Pb collisions at 2.76 TeV~\cite{aliceds} as shown in the right panel of
Fig.~\ref{ds_by_d0}. Although collision centrality and $p_{T}$ range are different, both the measurements are qualitatively consistent within the
large uncertainties.\\
The nuclear modification factor, $R_{\rm{AA}}$, is
defined as the ratio of the particle yield in the Au+Au collision to
$p+p$ collisions normalized by the number of binary
collisions ($N_{\rm{bin}}$). The value of  $N_{\rm{bin}}$ is calculated from a
Monte Carlo Glauber simulation. The measured charm cross-section in $p+p$ collisions at $\sqrt{s}$ = 200 GeV~\cite{stard0pp} is used to get the $D_{S}$ spectra for $p+p$ collisions.
The fragmentation factor ($c$ $\longrightarrow$ $D_{S}$) used to convert the charm
cross-section to $D_{S}$ production cross-section is
0.09 $\pm$ 0.01~\cite{frag0,frag1,frag2}. The Levy fit function is used
to get $D_{S}$ yield for $p+p$ collisions (as shown in the left panel of  Fig.~\ref{raa}) at corresponding measured
$p_{T}$  in Au+Au collisions. 
The $R_{\rm{AA}}$ of  $D_{S}$ at mid-rapidity ($|y| < 1.0$) in
Au+Au collisions at $\sqrt{s_{NN}}$ = 200 GeV for 10-40$\%$ centrality
is presented in the right panel of Fig.~\ref{raa}.
The $R_{\rm{AA}}$ of $D_{S}$  is equal to 2.1 $\pm$ 0.5 (stat.) $\pm$$^{0.7}_{0.7}$ (sys.) and
1.7 $\pm$ 0.4 (stat.) $\pm$$^{0.5}_{0.7}$ (sys.) at $p_{T}$ = 2.8 and 3.9
GeV/$c$, respectively. The  $R_{\rm{AA}}$ of $D_{S}$  is
systematically higher than that of  $D^{0}$~\cite{stard0}  although the
enhancement is not statistically significant. \\
The elliptic flow ($v_{2}$), a measure of the anisotropy in the momentum
space,  can be used to probe the dynamics of early stages of heavy-ion collisions~\cite{hydro}.
The full event plane method~\cite{method} has been used for the flow                                                                                       
analysis, where the event plane is calculated using measured charged                                                                                       
particle in  the TPC at mid-pseudorapidity ($|\eta|<1.0$). Event plane
resolution has been calculated using $\eta$ sub-event plane method.
Fig.~\ref{v2_fig} shows  $v_{2}$ as a function of
$p_{T}$ for $D_{S}$, $D^{0}$~\cite{d0v2} and $\phi$~\cite{200GeV_run11_star} for 0-80$\%$ centrality in Au + Au collisions at
$\sqrt{s_{NN}}$ = 200 GeV. The statistical error on $D_{S}$ $v_{2}$ is
very large, therefore it is not possible to make any conclusion from
the current study. We will collect more data in the year 2016 with improved HFT performance~\cite{contin}, which will allow us to measure $D_{S}$ $v_{2}$ with reasonable statistical error.
\section{Summary}
\label{summary}
We report the first measurements of $D_{S}$-meson production and elliptic flow at mid-rapidity in Au + Au
collisions at $\sqrt{s_{NN}}$ = 200 GeV recorded in 2014 with the STAR experiment.
At the intermediate $p_{T}$, the ratio $D_{S}/D^{0}$ seems to be
higher than the prediction in $p+p$ collisions by PYTHIA. The nuclear modification factor of $D_{S}$ are
systematically higher than unity and $D^{0}$ $R_{\rm{AA}}$. Our study
is limited by statistics but indicates a hint of enhancement of
$D_{S}$ production in Au+Au collisions with respect to $p+p$ collisions
as compared to non-strange $D$ mesons. We have also presented the
first measurement of $D_{S}$ $v_{2}$. High statistics data, which will
be collected in 2016 by STAR, will be helpful to study the $D_{S}$ production
mechanism and  $v_{2}$ in Au+Au collisions.


\bibliographystyle{elsarticle-num}
\bibliography{<your-bib-database>}

\end{document}